\documentclass[aps,prl,reprint,groupedaddress]{revtex4-1}

\usepackage{color}
\usepackage{graphicx}
\usepackage{amsmath}
\usepackage{bm}
\usepackage{amsfonts}

\newcommand{\leftp}{\left(}
\newcommand{\rightp}{\right)}

\begin{document}
\raggedbottom


\title{Wireless Josephson Amplifier}


\author{A. Narla, K.M. Sliwa, M. Hatridge, S. Shankar, L. Frunzio, R. J. Schoelkopf, M.H. Devoret}
\affiliation{Department of Applied Physics, Yale University, New Haven, Connecticut 06511, USA}


\date{\today}

\begin{abstract}
Josephson junction parametric amplifiers are playing a crucial role in the readout chain in superconducting quantum information experiments. However, their integration with current 3D cavity implementations poses the problem of transitioning between waveguide, coax cables and planar circuits. Moreover, Josephson amplifiers require auxiliary microwave components, like directional couplers and/or hybrids, that are sources of spurious losses and impedance mismatches that limit measurement efficiency and amplifier tunability. We have developed a wireless architecture for these parametric amplifiers that eliminates superfluous microwave components and interconnects. This greatly simplifies their assembly and integration into experiments. We present an experimental realization of such a device operating in the $9-11$~GHz band with about $100$~MHz of amplitude gain-bandwidth product, on par with devices mounted in conventional sample holders. The simpler impedance environment presented to the amplifier also results in increased amplifier tunability.
\end{abstract}

\maketitle

Parametric amplifiers (paramps) based on Josephson junctions \cite{Wahlsten1978, Mygind1979, Yurke1989}, such as the Josephson Bifurcation Amplifier (JBA) and its degenerate amplifier relatives\cite{Siddiqi2004, Siddiqi2005, Castellanos-Beltran2007, Vijay2008, Yamamoto2008, Castellanos-Beltran2008, Kamal2009, Hatridge2011, Mutus2013, Zhong2013} and the Josephson Parametric Converter (JPC)\cite{Bergeal2010, Bergeal2010a, Abdo2011, Roch2012, Abdo2013, Schackert2013}, have become essential components in the quantum non-demolition (QND) readout chain of superconducting qubits. They have been instrumental in the experimental observation of quantum jumps\cite{Vijay2011}, the detailed study of the quantum backaction of measurement\cite{Hatridge2013, Murch2013}, and feedback\cite{Vijay2012n, deLange2013}, and will remain essential in future experiments aimed at quantum error correction and beyond\cite{Sun2013, Devoret2013}. However, their integration into current state-of-the-art 3D circuit quantum electrodynamics (cQED) experiments\cite{Paik2011, Rigetti2012} introduces the problem of interconnects that transition between waveguide, coaxial and planar microwave environments. Furthermore, these amplifiers require auxiliary microwave components such as directional couplers and/or hybrids to operate.  These components introduce two problem: i) losses that reduce the system measurement efficiency and thus the fidelity of qubit readout; ii) a complicated frequency dependence of the impedance seen by the device which limits amplifier tunability and performance.

We radically simplify the design and implementation of an amplifier such as the JBA by coupling the lumped element Josephson circuit of the amplifier directly to the propagating mode of a rectangular waveguide using a dipole antenna. This design uses only low-loss, high quality materials commonly found in 3D superconducting qubits and eliminates printed circuit (PC) boards and wirebonds, as well as most of the auxiliary microwave components and interconnects. Their removal greatly simplifies the impedance seen by the device, making it flatter as a function of frequency, thus leading to improved amplifier tunability. 

\begin{figure*}
\includegraphics{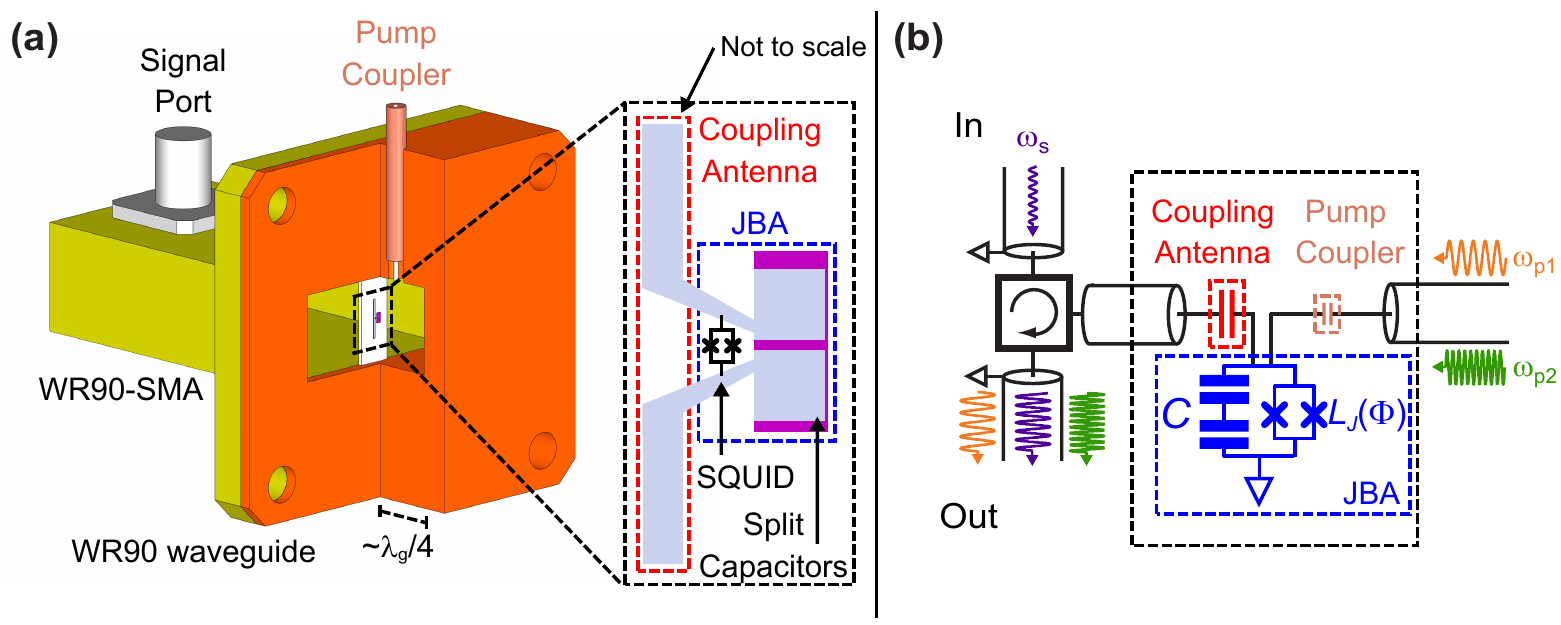}
\caption{\textbf{Figure 1 | Wireless Josephson Amplifier Schematic.} 
A) Wireless Architecture for Josephson Amplifiers: The WJA circuit consists of a SQUID  shunted by a split parallel-plate capacitor (blue box) fabricated on a sapphire chip. The tunable inductance of the SQUID is $L_J(\Phi) = (\Phi_0/2\pi I_0)$cos$(\pi\Phi/\Phi_0)$ where $I_0$ is the SQUID critical current, $\Phi$ is a magnetic flux applied through an external coil and $\Phi_0$ is the magnetic flux quantum. This lumped-element circuit is connected to an antenna (red box) that wirelessly couples the circuit to propagating electromagnetic waves. This chip is placed inside a copper WR-90 rectangular waveguide at a distance of $\lambda_g/4$ away from a wall (effectively a short) that reflects all traveling waves, situating it at an anti-node of the standing wave. The device amplifies in reflection with all signals at the operating frequency $\omega_s$ entering and exiting the waveguide port. For this experiment, a WR90-coax adapter is used to integrate the experiment with existing fridge wiring.  B) Circuit schematic: The circuit equivalent of the schematic in (a) is shown in the dashed black box. Small signals enter the device via the WR90-coax adapter and travel down the waveguide before exciting a differential signal across the resonator via the antenna. On the other hand, the large pump tones at $\omega_{p1}$ and $\omega_{p2}$ required for amplification enters the circuit through a weakly coupled pump port. The process of four-wave mixing in the non-linear resonator causes the small signal at $\omega_s$ to be amplified and then re-radiated into the waveguide by the coupling antenna along with the pump tones.}
\end{figure*}

In this article, we describe the device, called the Wireless Josephson Amplifier (WJA). It is similar to the conventional JBA \cite{Hatridge2011, Vijay2008} in that it  consists of a lumped element resonator made of a superconducting quantum interference device (SQUID), acting as an external flux-tunable inductance $L_J(\Phi) = (\Phi_0/2\pi I_0)$cos$(\pi\Phi/\Phi_0)$, shunted by a split parallel-plate capacitor with capacitance $C$ (see Fig.~1). However, unlike the conventional implementation, a dipole antenna galvanically connected to the resonator couples it directly to the lowest-order propagating electromagnetic mode in a rectangular waveguide, which is chosen to be WR-90 (with inner dimensions of 0.90 inches by 0.40 inches) so that the WJA operates between $8.2$ and $12.4$~GHz. The chip is placed a quarter wavelength away from a wall - effectively a shorted termination - situating it at an electric field anti-node. Signals at frequency $\omega_s$ enter the device through a waveguide to coaxial cable (coax) adapter (used to make the device compatible with the existing experimental setup but will be eliminated in future experiments), traveling down the waveguide before exciting a differential signal across the resonator. This differential excitation makes the hybrids of the conventional implementation unnecessary. The pump tones at $\omega_{p1}$ and $\omega_{p2}$ required for stiff-pump amplification\cite{Kamal2009} enter through a separate weakly coupled pump port, thereby eliminating the need for a directional coupler. Four-wave mixing in the non-linear resonator results in the signals at $\omega_s$ being amplified and re-radiated by the coupling antenna. The short ensures that all signals preferentially exit the device through the waveguide port.

Aside from the simplified design, the effective circuit representation of the WJA remains unchanged from the conventional implementation\cite{Hatridge2011, Vijay2008} except for one crucial aspect: the added dipole antenna to the WJA behaves like a coupling capacitor ($C_c$) between the resonator and the waveguide (see Fig.~1(b)). As a result, unlike the conventional JBA, the resonator bandwidth can be engineered independently of the resonator frequency $\omega_0 = 1/\sqrt{(L_J+L_{stray})C}$ because the coupling $Q$ of the resonator is given by:
\begin{equation}
Q = \frac{Z_c}{Z_L(d)}\leftp\frac{C}{C_c}\rightp^2
\end{equation}
where $Z_L(d)$ is the impedance seen by the resonator which is a function of the chip distance from the wall, $d$, $Z_c = \sqrt{(L_J+L_{stray})/C}$ is the characteristic impedance of the resonator, where $L_{stray}$ is the stray inductance in the circuit. Thus, the resonator $Q$ can be tuned controllably by changing $C_c$, which in turn is proportional to the length of the dipole antenna, $l$, in the limit of short lengths. To realize an amplifier designed for qubit readout, the circuit parameters were chosen to be $I_0 = 4~\mu$A and $C=3.5$~pF for a maximum linear resonance frequency around $9.5$~GHz; in addition, to make the amplifier bandwidth larger than the typical qubit cavity bandwidth, we chose $Q=100$ which required two antenna pads that were each $2.5$~mm long and $0.25$~mm wide separated by $150~\mu$m gaps, sizes that can be readily fabricated using either optical or electron-beam lithography.

Although $Q=100$ was chosen for this particular device, as shown in Fig.~2(a), the $Q$ can be tuned by over three orders of magnitude by choosing $l$ between $0.5$~mm and $5$~mm. While the $Q(l)$ dependence can be calculated analytically, a finite element electromagnetic solver, such as HFSS\footnote{ANSYS HFSS version 15.1}, was used for a more detailed analysis. These simulations confirmed that the $Q\propto l^{-2}$ behavior predicted by Eqn.~1 for short antenna lengths breaks down for longer antenna lengths when the size of the dipole antenna becomes comparable to the size of the waveguide. The behavior of the coupling as a function of the chip distance from the wall, $d$, was also simulated; the coupling should be maximum when the chip is at an antinode of the electric field in the waveguide and minimum when it is at a node. As shown in Fig.~2(b), the simulations confirm the expected divergence in the $Q$ when $d=n\lambda_{g}/2$ where $n$ is an integer and $\lambda_g$ is the guide wavelength\cite{Pozar2012}; the $Q$ diverges at these points because $Z_L(d=n\lambda_{g}/2)=0$. Moreover, because the coupling is a flat function of $d$ (to first order) around $d=(2n+1)\lambda_{g}/4$, the design is robust against errors in fabrication or machining that affect the resonator frequency or its position from the wall.

\begin{figure*}
\includegraphics{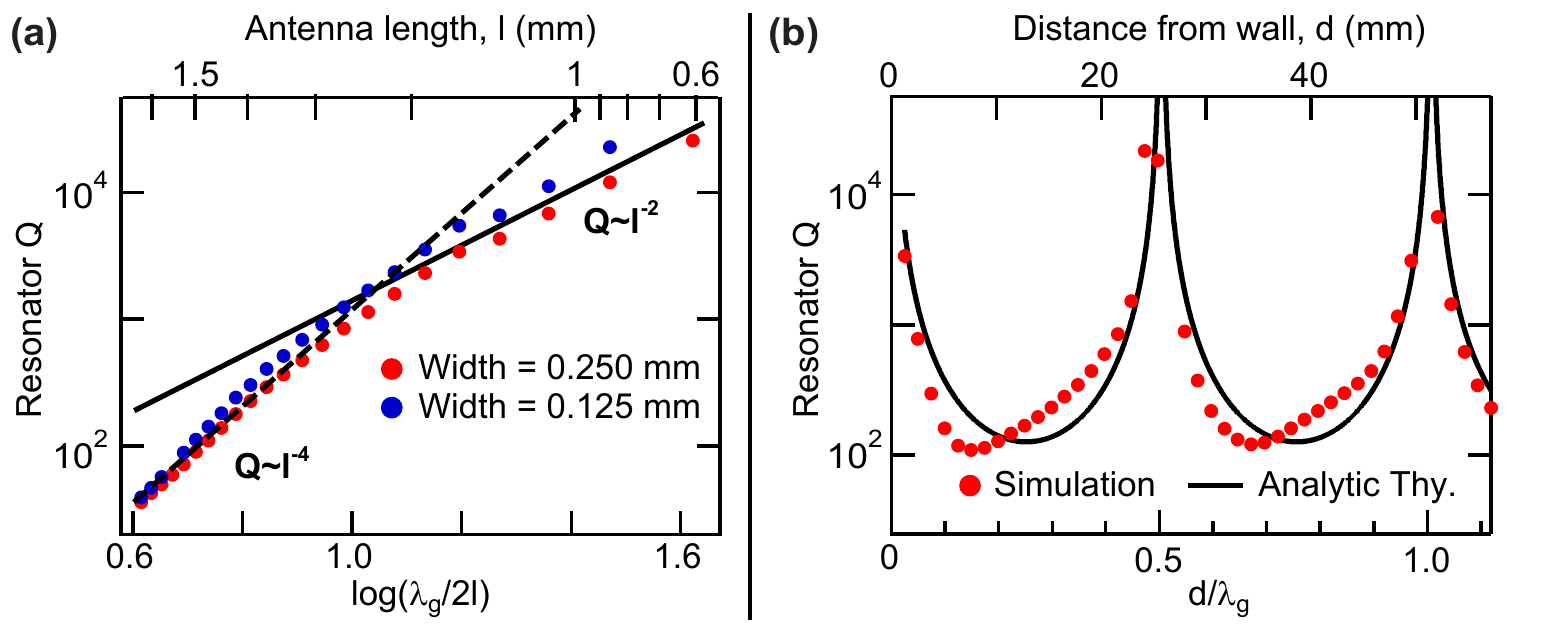}
\caption{\textbf{Figure 2 | Predicted Device Coupling.}
(a) Linear Q versus Antenna Length: Simulations of the linear quality factor of the WJA resonator as a function of the antenna length confirm that the $Q\approx l^{-2}$ for small antenna lengths as would be expected for a dipole antenna. This dependence breaks down for longer lengths as the antenna becomes comparable in size to the waveguide. Nevertheless, the Q can be simulated and tuned over three orders of magnitude by changing the length of the antenna. Changing the antenna width only leads to small changes in the coupling Q as expected. (b) Linear Q versus Chip Position: The linear Q of the WJA circuit is minimum when the chip is placed at $\lambda_g/4$ away from the wall of the waveguide indicating maximum coupling. This Q diverges at $n\lambda_g/2$ since the chip is at a node of the standing wave. Because the Q around $\lambda_g/4$ only slowly varies with the distance between the chip and the wall, the device is robust to errors in either its frequency or its placement in the waveguide.}
\end{figure*}

Fabricating the lumped-element circuit for the WJA required making both parallel-plate capacitors and a Josephson junction on chip. The same three-layer fabrication process developed for conventional JBAs\cite{Frunzio2006, Vijay2008, Hatridge2011} was used to avoid the complication of making vias. First, a $150$~nm niobium layer was deposited on the wafer and patterned using optical lithography and reactive-ion etching to define the shared ground plane of the split parallel-plate capacitors. Next, $200$~nm silicon nitride was deposited conformally over the entire wafer using plasma-enhanced chemical vapor deposition to form the dielectric layer of the capacitors; the thickness and dielectric constant ($\epsilon_r=7$) of this layer are used to calculate the size of the capacitors needed for a $3.5$~pF capacitance. Finally, a bi-layer of electron-beam resist was spun on the wafer and patterned using electron-beam lithography. Double angle evaporation of aluminum was used to define the Josephson junctions in the $8~\mu$m $\times$ $2~\mu$m SQUID loop as well as the top plates of the capacitors and the coupling antenna.

\begin{figure}
\includegraphics{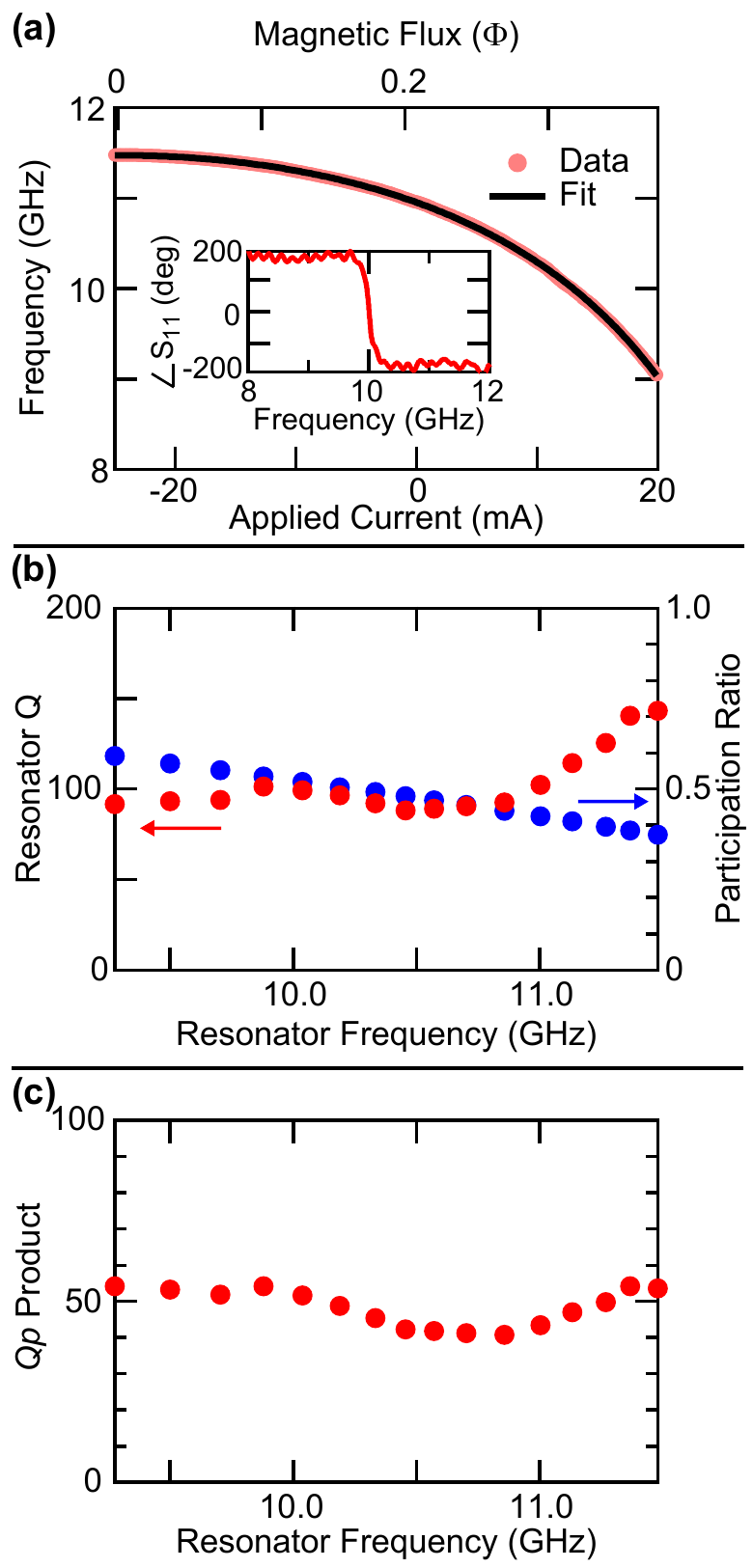}
\caption{\textbf{Figure 3 | Q and Participation Ratio versus Resonator Frequency.} 
(a) Resonance Frequency versus Flux: The reflected phase (inset) is plotted as a function of frequency revealing a $360^{\circ}$ degree phase roll characteristic of an over-coupled resonator. Changing the flux allows this resonance frequency to be tuned over $2$~GHz. By fitting the resonance frequency as a function of the flux through the SQUID, the circuit parameters of the device are extracted to be $I_0=4.6\mu$A, $C=1$~pF and $L_{stray}=120$~pH. B) $Q$ and $p$ versus Frequency: A plot of the resonator $Q$ as a function of the resonance frequency shows that $Q\approx100$, varying by around $10\%$ over most of the frequency range except above $11$~GHz where it rapidly increases by around $40\%$; this is because at these higher frequencies, the chip location is significantly different from the electric field anti-node. On the other hand, the participation ratio varies smoothly with resonance frequency; at the highest resonator frequency, the participation ratio is the smallest since $L_J$ is at its maximal value. (b) $Qp$ Product versus Frequency: Over the range of operation, the $Qp$ product is relatively flat varying by about $20\%$ above and below its average value of $Qp=50$.}
\end{figure}

The WJAs were cooled down in a cryogen-free dilution refrigerator to about $50$~mK. First, the amplifier's linear resonance frequency and bandwidth were measured by looking at the reflected phase using a vector network analyzer (VNA). The reflected phase (Fig.~3(a) inset) showed a $360^\circ$ phase roll characteristic of an overcoupled resonator. From this, the coupling was found to be $Q=100$ as desired and agreeing with simulation results. Next, by applying current to an external superconducting coil magnet, the linear frequency of the WJA was tuned with flux. As shown in Fig.~3(a), the linear resonance frequency tuned from its maximum value of $11.4$~GHz down to $9$~GHz. By fitting the resonance frequency as a function of flux to the equation:
\begin{equation}
f(\Phi) = \frac{1}{2\pi\sqrt{CL_J\leftp \frac{L_{stray}}{L_J}+\frac{1}{\cos{\pi\Phi/\Phi_0}}\rightp}} \label{fJBA}
\end{equation}
the circuit parameters were found to be  $C = 1$~pF, $L_{stray}=120$~pH and $I_0=4.6~\mu$A with error bars of about $10\%$. The discrepancy in the measured capacitance from the desired value is a result of operating the circuit near the self-resonance frequency of the capacitor; in future iterations of the WJA, it would be advantageous to use thinner dielectric layers and/or larger dielectric constants to avoid this problem. Nevertheless, the device still operated in the desired frequency range. More importantly, as shown in Fig.~3(b), the $Q$ varied by only $10\%$ between $9$~GHz and $11$~GHz. Above $11$~GHz, the $Q$ changed rapidly because at these higher frequencies, the chip distance from the wall is significantly different from $\lambda_g/4$. The almost constant $Q$ as a function of frequency not only agrees well with simulation results but also suggests that the environment's impedance is well matched over the whole frequency range. Because the $Q$ of the WJA can be varied by changing the antenna length independent of the device frequency, this architecture offers a valuable control knob over the amplifier bandwidth that is unavailable in conventional JBAs.

Another important figure of merit for single effective-junction parametric amplifiers is the inductance participation ratio, $p=L_J/(L_J+L_{stray})$. Using the the extracted values of $L_{stray}$ and $L_J$, the $p$ was found to vary between about $0.6$ and $0.4$ over the entire frequency range (see blue trace in Fig.~3(b)). Together the $Q$ and $p$ are important because their product determines both whether the resonator will amplify and the maximum signal power that can be amplified without saturation\cite{Schackert2013}. To realize an amplifier with sufficient gain for qubit readout, it is necessary that $Qp\gtrsim5$; however, to maximize the amplifier saturation power, the $Qp$-product needs to be as low as possible with $Qp\approx 5$ to $10$ being optimal\cite{Schackert2013}. In particular, for the present device, the $Qp$ product was around $50$ by design. Since $Qp$ varies by only about $20\%$ over the whole frequency range (see Fig.~3(c)), this device is expected to show tunability and constant amplifier performance over the operation band. Combined with the increased control over the $Q$ and $p$ in this architecture, this will allow for fuller optimization of the amplifier's properties in future iterations.

The WJA was characterized as a phase-sensitive amplifier by applying two pump tones, $\omega_{p1}$ and $\omega_{p2}$, symmetrically detuned by $500$~MHz from the signal tone at $\omega_s$ through the weakly coupled pump port. Different gains at the same frequency were achieved by keeping the flux through the SQUID constant while changing the two pump powers. As shown in Fig.~4(a), the amplifier  achieved gains between unity and over $25$~dB. In particular at $20$~dB of gain, the instantaneous amplifier bandwidth is $13$~MHz (much larger than typical cavity bandwidths used in qubit experiments). 

The amplifier's dynamic range was characterized by measuring its maximum gain as a function of input signal power from the VNA and finding the power at which the gain fell by $1$~dB (the $P_{-1dB}$ power); at $20$~dB of gain, we found that the saturation power was $-132$~dBm which corresponded to approximately $0.7$ photons in the $13$~MHz amplifier bandwidth (see Fig.~4(b)). Additionally, the slope of $P_{-1dB}$ power versus gain (orange line in Fig.~4(b)) was $-1.2$~dB/dB, close to the ideal slope of $-1$~dB/dB expected for an ideal linear parametric amplifier\cite{Abdo2011}. The dynamic range of this device, while comparable to conventional JBAs, could be further increased by reducing the $Qp$-product by a factor of ten to $Qp\approx 5$; while this could be done either by reducing the $Q$ or the $p$, the current $Q\approx100$ is optimal for qubit readout and so reducing the participation ratio is desirable. This can be achieved, for example, by using multiple SQUIDs with larger $I_0$ and hence a smaller $L_J$.

\begin{figure}
\includegraphics{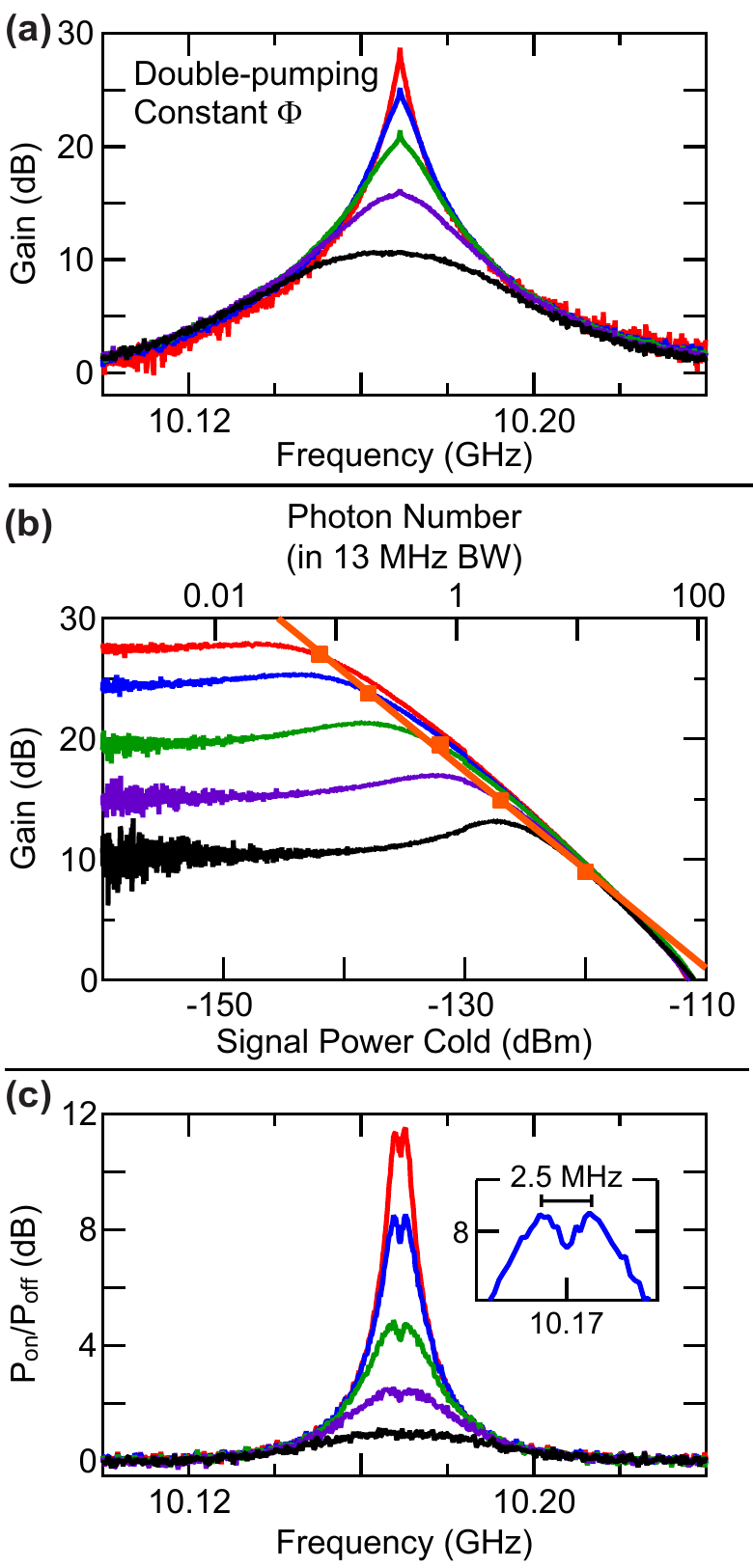}
\caption{\textbf{Figure 4 | Amplifier Performance.} 
(a) Gain:  The WJA is pumped with two pump tones that are symmetrically detuned by $500$~MHz from the center frequency at fixed flux. Gains (acquired at $-140$~dBm signal power except for the $28$~dB gain curve taken at $-150$~dBm) up to $28$~dB (red trace) can be achieved with this device; the pump powers required vary between $-64.51$~dBm and $-63.42$~dBm for $28$~dB (red trace) gain to $-65.43$~dBm and $-64.14$~dBm for $10$~dB (black trace) gain. At $20$~dB (green trace) of gain, the amplifier has an instantaneous bandwidth of $13$~MHz. (b) Dynamic Range: At $20$~dB of gain, the input signal power required to saturate the amplifier and reduce its gain by $1$~dB ($P_{-1dB}$) is about $0.7$ photons in the instantaneous resonator bandwidth of $13$~MHz. The slope of the $P_{-1dB}$ powers versus gain is $-1.2$~dB/dB.  (c) Noise Visibility Ratio: The increase in noise measured at room temperature is plotted for different amplifier gains; at a relatively low gain of $10$~dB the WJA does not amplify quantum noise above the noise added by the following HEMT amplifier. However, when the gain is increased, the amplified quantum noise becomes larger than the added noise from the HEMT. The dips in the traces (inset) at the center frequency are because the WJA behaves as a phase-sensitive amplifier at $\omega_s$.}
\end{figure}

A final figure of merit, especially for parametric amplifiers, is the amplifier's noise temperature, $T_N$. According to Caves\cite{Caves1982}, when the WJA is operated as a phase-sensitive parametric amplifier, the added noise is zero for the amplified quadrature ($T_N=0$) while $T_N=\infty$ for the de-amplified quadrature.  The noise temperature for this amplifier was estimated from the increase in the noise measured by a room temperature spectrum analyzer when the WJA was turned on. When the WJA was off, i.e $G=0$, the noise measured at room temperature is entirely dominated by noise added by the high electron mobility transistor (HEMT) amplifier. On the other hand, when the WJA is turned on, the measured noise increases because the spectrum analyzer now also receives amplified quantum noise from the parametric amplifier. We call the noise visibility ratio the amount by which the combined noise of the WJA and the HEMT is greater than the HEMT noise alone.

As shown in Fig.~4(c), the noise visibility increased with WJA gain; when $G=25$~dB (blue trace), the ratio was about $8.5$~dB indicating that less than $15\%$ of the power at room temperature was added HEMT noise. The dips in the traces at the center frequency are a consequence of the WJA behaving as a phase-sensitive amplifier at $\omega_s$. Since only one of the two quadratures is amplified, the measured noise should be $3$~dB lower. However, the finite resolution bandwidth of the spectrum analyzer smears out the dip and consequently, the measured dips were smaller than the expected $3$~dB (see inset of Fig.~4(c)) and have a width of $2.5$~MHz equal to the spectrum analyzer's resolution bandwidth. At $G=20$~dB, the difference in noise power of $4.5$~dB was comparable to other parametric amplifiers, JBAs or JPCs, measured in this experimental setup indicating that the noise performance of the WJA is comparable to standard paramps. For this measurement setup, the system noise temperature, $T_{sys}^{H}$, which is the effective noise temperature of the HEMT and the losses between the plane of the WJA and the HEMT, is assumed to be $35$~K. From the measured increase of $4.5$~dB at $G=20$~dB, we estimate an upper bound on the noise added by the WJA of $T_N\leq2.5T_Q$, where $T_Q = \hbar\omega_s/2k_B$. 

In conclusion, we have designed and operated a wireless implementation of a Josephson parametric amplifier for readout of a superconducting qubit in a 3D cavity. Our design replaces the usual connections between the chip of the amplifier and the qubit port by a simple on-chip antenna placed in the center of a waveguide. This leads to a much simplified assembly of critical components. The WJA exhibits the same gain, bandwidth, dynamic range and noise properties as a conventional JBA while also offering improved tunability and increased control over both the quality factor $Q$ and participation ratio $p$.  With this better control, it is possible to increase the amplifier dynamic range with a WJA consisting of multiple SQUIDs in series with correspondingly large critical current junctions. The simplification of the microwave environment seen by the Josephson element should also eliminate sources of loss that currently limit the measurement efficiency of circuit QED experiments. The incorporation of a WJA in a qubit readout chain is currently under way. Finally, this architecture could be easily extended to realize a wireless JPC by adding dipole antenna to a lumped JPC circuit and placing it at the intersection of two waveguides with perpendicular polarizations.

\begin{acknowledgments}
Fabrication assistance from M. Rooks, as well as the Yale Institute for Nanoscience and Quantum Engineering facilities are gratefully acknowledged. This research was supported by the National Science Foundation under Grants No. DMR-1006060 and No. DMR- 0653377, the US Army Research Office Grants No. W911NF-09-1-0514 and No. W911NF-14-1-0011.
\end{acknowledgments}

\end{document}